%% file: main.tex
\pdfoutput=1

\documentclass[10pt,letter,prl,reprint,twocolumn,hypertext]{revtex4-1}

\usepackage{ifthen} 
\newboolean{pdflatex}
\setboolean{pdflatex}{true} 

\newboolean{articletitles}
\setboolean{articletitles}{true} 

\newboolean{uprightparticles}
\setboolean{uprightparticles}{false} 

\usepackage{microtype}
\usepackage{lineno}  
\usepackage{xspace} 
\usepackage{bigstrut}

\usepackage{graphicx}  
\usepackage{color}
\usepackage{colortbl}
\graphicspath{{./figs/}} 

\usepackage{amsmath} 
\usepackage{amssymb}
\usepackage{amsfonts}
\usepackage{upgreek} 

\newcommand*\patchAmsMathEnvironmentForLineno[1]{%
\expandafter\let\csname old#1\expandafter\endcsname\csname #1\endcsname
\expandafter\let\csname oldend#1\expandafter\endcsname\csname
end#1\endcsname
 \renewenvironment{#1}%
   {\linenomath\csname old#1\endcsname}%
   {\csname oldend#1\endcsname\endlinenomath}%
}
\newcommand*\patchBothAmsMathEnvironmentsForLineno[1]{%
  \patchAmsMathEnvironmentForLineno{#1}%
  \patchAmsMathEnvironmentForLineno{#1*}%
}
\AtBeginDocument{%
\patchBothAmsMathEnvironmentsForLineno{equation}%
\patchBothAmsMathEnvironmentsForLineno{align}%
\patchBothAmsMathEnvironmentsForLineno{flalign}%
\patchBothAmsMathEnvironmentsForLineno{alignat}%
\patchBothAmsMathEnvironmentsForLineno{gather}%
\patchBothAmsMathEnvironmentsForLineno{multline}%
}

\usepackage{hyperref}    
\usepackage[all]{hypcap} 

\input{lhcb-symbols-def} 

\usepackage{mciteplus}

\input{bhhh-symbols-def}  

\begin{document}



\title{Measurement of $C\!P$ violation in the phase space of\\ $B^{\pm} \to K^{\pm} \pi^{+} \pi^{-}$ and $B^{\pm} \to K^{\pm} K^{+} K^{-}$ decays \vspace{0.4cm}}

\vspace*{1cm}
\author{\input{LHCb_HD_authorlist}}
\collaboration{The LHCb collaboration\vspace{0.4cm}}

\begin{abstract}
The charmless decays $B^{\pm}\to K^{\pm}\pi^+\pi^-$ and $B^{\pm}\to K^{\pm}K^+K^-$ are reconstructed using data, corresponding to an integrated luminosity of 1.0~fb$^{-1}$, collected by LHCb in 2011. 
The inclusive charge asymmetries of these modes are measured as $ A_{C\!P}(B^{\pm}\to K^{\pm}\pi^+\pi^-) = 0.032 \pm 0.008 \stat \pm 0.004 \syst \pm 0.007 (J\mskip -3mu/\mskip -2mu\Ppsi\mskip 2mu K^{\pm})$ and $A_{C\!P}(B^{\pm}\to K^{\pm}K^+K^-)   =  -0.043 \pm 0.009 \stat \pm 0.003 \syst \pm 0.007 (J\mskip -3mu/\mskip -2mu\Ppsi\mskip 2mu K^{\pm})$, where the third uncertainty is due to the $C\!P$ asymmetry of the $B^{\pm}\to J\mskip -3mu/\mskip -2mu\Ppsi\mskip 2mu K^{\pm}$ reference mode. 
The significance of $A_{C\!P}(B^{\pm}\to K^{\pm}K^+K^-)$ exceeds three standard deviations and is the first evidence of an inclusive $C\!P$ asymmetry in charmless three-body $B$ decays.  
In addition to the inclusive $C\!P$ asymmetries, larger asymmetries are observed in localised regions of phase space. 
\end{abstract}

\pacs{Valid PACS appear here}

\vspace*{-1.5cm}
\hspace{-8.6cm}
\mbox{\Large EUROPEAN ORGANIZATION FOR NUCLEAR RESEARCH (CERN)}

\vspace*{0.7cm}
\hspace*{-9cm}
\begin{tabular*}{16.6cm}{lc@{\extracolsep{\fill}}r}
\ifthenelse{\boolean{pdflatex}}
{\vspace*{-3.2cm}\mbox{\!\!\!\includegraphics[width=.14\textwidth]{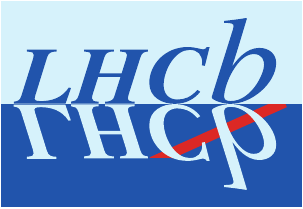}} & &}%
{\vspace*{-1.2cm}\mbox{\!\!\!\includegraphics[width=.12\textwidth]{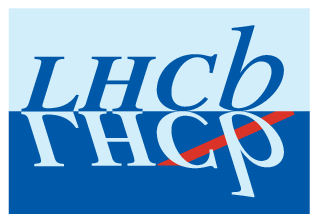}} & &}%
\\ 
 & &  \\  
 & &  \\  
 & &  \\  
 & &  \\  
 & & CERN-PH-EP-2013-090 \\  
 & & LHCb-PAPER-2013-027 \\  
 & & 5 June 2013 \\ 
 & & \\
\end{tabular*}

\vspace*{21.0cm}
\hspace*{-5cm}\centerline{\copyright~CERN on behalf of the LHCb collaboration, license \href{http://creativecommons.org/licenses/by/3.0/}{CC-BY-3.0}.}
\vspace*{-21.0cm}

\vspace*{20.0cm}
\hspace*{-5cm}\centerline{\large Submitted to Phys.~Rev.~Lett.}
\vspace*{-20.0cm}


\boldmath
\maketitle
\unboldmath




\input{body}

\input{acknowledgements}

\addcontentsline{toc}{section}{References}
\bibliographystyle{LHCb}
\bibliography{main,LHCb-PAPER,LHCb-CONF,LHCb-DP}

\end{document}

%% file: lhcb-symbols-def.tex
\def\lhcb {\mbox{LHCb}\xspace}
\def\ux85 {\mbox{UX85}\xspace}

\ifthenelse{\boolean{uprightparticles}}%
{

 \def\Pmu         {\ensuremath{\upmu}\xspace}

 \def\Ppi         {\ensuremath{\uppi}\xspace}

 \def\Ppsi        {\ensuremath{\uppsi}\xspace}

 \def\PDelta      {\ensuremath{\Delta}\xspace}                 
 \def\PXi      {\ensuremath{\Xi}\xspace}                 
 \def\PLambda      {\ensuremath{\Lambda}\xspace}                 
 \def\PSigma      {\ensuremath{\Sigma}\xspace}                 
 \def\POmega      {\ensuremath{\Omega}\xspace}                 
 \def\PUpsilon      {\ensuremath{\Upsilon}\xspace}                 
 
 \def\PB      {\ensuremath{\mathrm{B}}\xspace}                 
                  
 \def\PD      {\ensuremath{\mathrm{D}}\xspace}

 \def\PJ      {\ensuremath{\mathrm{J}}\xspace}                 
 \def\PK      {\ensuremath{\mathrm{K}}\xspace}

 \def\Pb      {\ensuremath{\mathrm{b}}\xspace}                 
 \def\Pc      {\ensuremath{\mathrm{c}}\xspace}

 \def\Pi      {\ensuremath{\mathrm{i}}\xspace}

 \def\Ps      {\ensuremath{\mathrm{s}}\xspace}

}
{

 \def\Pmu         {\ensuremath{\mu}\xspace}

 \def\Ppi         {\ensuremath{\pi}\xspace}

 \def\Ppsi        {\ensuremath{\psi}\xspace}                 
                  
 \mathchardef\PDelta="7101
 \mathchardef\PXi="7104
 \mathchardef\PLambda="7103
 \mathchardef\PSigma="7106
 \mathchardef\POmega="710A
 \mathchardef\PUpsilon="7107
                  
 \def\PB      {\ensuremath{B}\xspace}                 
                  
 \def\PD      {\ensuremath{D}\xspace}

 \def\PJ      {\ensuremath{J}\xspace}                 
 \def\PK      {\ensuremath{K}\xspace}

 \def\Pb      {\ensuremath{b}\xspace}                 
 \def\Pc      {\ensuremath{c}\xspace}

 \def\Pi      {\ensuremath{i}\xspace}

 \def\Ps      {\ensuremath{s}\xspace}

}

\def\mup        {\ensuremath{\Pmu^+}\xspace}

\def\squark    {\ensuremath{\Ps}\xspace}

\def\cquark    {\ensuremath{\Pc}\xspace}

\def\bquark    {\ensuremath{\Pb}\xspace}

\def\pion  {\ensuremath{\Ppi}\xspace}

\def\pip   {\ensuremath{\pion^+}\xspace}
\def\pim   {\ensuremath{\pion^-}\xspace}

\def\pipm  {\ensuremath{\pion^\pm}\xspace}
\def\pimp  {\ensuremath{\pion^\mp}\xspace}

\def\kaon  {\ensuremath{\PK}\xspace}
  \def\Kbar  {\kern 0.2em\overline{\kern -0.2em \PK}{}\xspace}

\def\Kz    {\ensuremath{\kaon^0}\xspace}
\def\Kzb   {\ensuremath{\Kbar^0}\xspace}
\def\KzKzb {\ensuremath{\Kz \kern -0.16em \Kzb}\xspace}
\def\Kp    {\ensuremath{\kaon^+}\xspace}
\def\Km    {\ensuremath{\kaon^-}\xspace}
\def\Kpm   {\ensuremath{\kaon^\pm}\xspace}
\def\Kmp   {\ensuremath{\kaon^\mp}\xspace}
\def\KpKm  {\ensuremath{\Kp \kern -0.16em \Km}\xspace}

\def\Kstarz  {\ensuremath{\kaon^{*0}}\xspace}

  \def\Dbar    {\kern 0.2em\overline{\kern -0.2em \PD}{}\xspace}
\def\D       {\ensuremath{\PD}\xspace}

\def\Dz      {\ensuremath{\D^0}\xspace}
\def\Dzb     {\ensuremath{\Dbar^0}\xspace}
\def\DzDzb   {\ensuremath{\Dz {\kern -0.16em \Dzb}}\xspace}
\def\Dp      {\ensuremath{\D^+}\xspace}
\def\Dm      {\ensuremath{\D^-}\xspace}

\def\DpDm    {\ensuremath{\Dp {\kern -0.16em \Dm}}\xspace}

\def\B       {\ensuremath{\PB}\xspace}
\def\Bbar    {\ensuremath{\kern 0.18em\overline{\kern -0.18em \PB}{}}\xspace}

\def\Bz      {\ensuremath{\B^0}\xspace}

\def\Bu      {\ensuremath{\B^+}\xspace}
\def\Bub     {\ensuremath{\B^-}\xspace}
\def\Bp      {\ensuremath{\Bu}\xspace}
\def\Bm      {\ensuremath{\Bub}\xspace}
\def\Bpm     {\ensuremath{\B^\pm}\xspace}

\def\Bs      {\ensuremath{\B^0_\squark}\xspace}

\def\jpsi     {\ensuremath{{\PJ\mskip -3mu/\mskip -2mu\Ppsi\mskip 2mu}}\xspace}

  \def\Y#1S{\ensuremath{\PUpsilon{(#1S)}}\xspace}

\def\Lbar {\ensuremath{\kern 0.1em\overline{\kern -0.1em\PLambda}}\xspace}

\def\to                 {\ensuremath{\rightarrow}\xspace}

\def\CP                {\ensuremath{C\!P}\xspace}
\def\CPT               {\ensuremath{C\!PT}\xspace}

\def\FL       {\ensuremath{F_{\mathrm{L}}}\xspace}
\def\AT#1     {\ensuremath{A_{\mathrm{T}}^{#1}}\xspace}           

\def\C#1      {\ensuremath{\mathcal{C}_{#1}}\xspace}                       
\def\Cp#1     {\ensuremath{\mathcal{C}_{#1}^{'}}\xspace}                    
\def\Ceff#1   {\ensuremath{\mathcal{C}_{#1}^{\mathrm{(eff)}}}\xspace}        
\def\Cpeff#1  {\ensuremath{\mathcal{C}_{#1}^{'\mathrm{(eff)}}}\xspace}       
\def\Ope#1    {\ensuremath{\mathcal{O}_{#1}}\xspace}                       
\def\Opep#1   {\ensuremath{\mathcal{O}_{#1}^{'}}\xspace}                    

\newcommand{\tev}{\ifthenelse{\boolean{inbibliography}}{\ensuremath{~T\kern -0.05em eV}\xspace}{\ensuremath{\mathrm{\,Te\kern -0.1em V}}\xspace}}
\newcommand{\gev}{\ensuremath{\mathrm{\,Ge\kern -0.1em V}}\xspace}
\newcommand{\mev}{\ensuremath{\mathrm{\,Me\kern -0.1em V}}\xspace}
\newcommand{\kev}{\ensuremath{\mathrm{\,ke\kern -0.1em V}}\xspace}
\newcommand{\ev}{\ensuremath{\mathrm{\,e\kern -0.1em V}}\xspace}
\newcommand{\gevc}{\ensuremath{{\mathrm{\,Ge\kern -0.1em V\!/}c}}\xspace}
\newcommand{\mevc}{\ensuremath{{\mathrm{\,Me\kern -0.1em V\!/}c}}\xspace}
\newcommand{\gevcc}{\ensuremath{{\mathrm{\,Ge\kern -0.1em V\!/}c^2}}\xspace}
\newcommand{\gevgevcccc}{\ensuremath{{\mathrm{\,Ge\kern -0.1em V^2\!/}c^4}}\xspace}
\newcommand{\mevcc}{\ensuremath{{\mathrm{\,Me\kern -0.1em V\!/}c^2}}\xspace}

\newcommand{\stat}{\ensuremath{\mathrm{\,(stat)}}\xspace}
\newcommand{\syst}{\ensuremath{\mathrm{\,(syst)}}\xspace}

\newcommand{\chisq}{\ensuremath{\chi^2}\xspace}

\def\gsim{{~\raise.15em\hbox{$>$}\kern-.85em
          \lower.35em\hbox{$\sim$}~}\xspace}
\def\lsim{{~\raise.15em\hbox{$<$}\kern-.85em
          \lower.35em\hbox{$\sim$}~}\xspace}

\def\pt         {\mbox{$p_{\rm T}$}\xspace}

\def\evtgen     {\mbox{\textsc{EvtGen}}\xspace}

\def\geant      {\mbox{\textsc{Geant4}}\xspace}

\def\photos     {\mbox{\textsc{Photos}}\xspace}

\def\pythia     {\mbox{\textsc{Pythia}}\xspace}

\def\tell1  {TELL1\xspace}
\def\ukl1   {UKL1\xspace}

\newcommand{\eg}{\mbox{\itshape e.g.}\xspace}

%% file: bhhh-symbols-def.tex
\def\pipipi {\ensuremath{{\Bpm \to \pip \pim \pipm}}\xspace}
\def\kpipi {\ensuremath{{\Bpm \to \Kpm \pip \pim}}\xspace}
\def\kkpi {\ensuremath{{\Bpm \to \Kp \Km \pipm}}\xspace}
\def\kkk {\ensuremath{{\Bpm \to \Kpm \Kp \Km}}\xspace}

\def\jpsik {\ensuremath{{\Bpm \to \jpsi \Kpm}}\xspace}

\def\kpipip {\ensuremath{{\Bp \to \Kp \pip \pim}}\xspace}
\def\kpipim {\ensuremath{{\Bm \to \Km \pip \pim}}\xspace}
\def\kkkp {\ensuremath{{\Bp \to \Kp \Kp \Km}}\xspace}
\def\kkkm {\ensuremath{{\Bm \to \Km \Kp \Km}}\xspace}

\def\kpipis {\ensuremath{{\Kpm \pip \pim}}\xspace}

\def\kkks {\ensuremath{{\Kpm \Kp \Km}}\xspace}

\def\mmkpi {\ensuremath{m^2_{\Kpm \pimp}}\xspace}                    
\def\mpipi {\ensuremath{m_{\pi \pi}}\xspace}                        
\def\mmpipi {\ensuremath{m^2_{\pip \pim}}\xspace}                        
\def\mkk {\ensuremath{m_{K K}}\xspace}                            
\def\mkpi {\ensuremath{m_{K \pi}}\xspace}                            
\def\mmkklow {\ensuremath{m^2_{\Kp \Km\,{\rm low}}}\xspace}     
\def\mmkkhi {\ensuremath{m^2_{\Kp \Km\,{\rm high}}}\xspace}      

\def\acp {\ensuremath{A_{\CP}}\xspace}
\def\acpraw {\ensuremath{A_{\rm raw}}\xspace} 

\def\acpn {\ensuremath{A_{\CP}^{N}}\xspace}

\def\adetk {\ensuremath{A_{\rm D}(\Kpm)}\xspace}
\def\aprod {\ensuremath{A_{\rm P}(\Bpm)}\xspace}


%% file: LHCb_HD_authorlist.tex
\centerline{\large\bf LHCb collaboration}
\begin{flushleft}
\small
R.~Aaij$^{40}$, 
B.~Adeva$^{36}$, 
M.~Adinolfi$^{45}$, 
C.~Adrover$^{6}$, 
A.~Affolder$^{51}$, 
Z.~Ajaltouni$^{5}$, 
J.~Albrecht$^{9}$, 
F.~Alessio$^{37}$, 
M.~Alexander$^{50}$, 
S.~Ali$^{40}$
% ...

{\footnotesize \it
% ...
$ ^{5}$Clermont Universit\'{e}, Universit\'{e} Blaise Pascal, CNRS/IN2P3, LPC, Clermont-Ferrand, France\\
$ ^{6}$CPPM, Aix-Marseille Universit\'{e}, CNRS/IN2P3, Marseille, France\\
% ...
$ ^{9}$Fakult\"{a}t Physik, Technische Universit\"{a}t Dortmund, Dortmund, Germany\\
% ...
$ ^{36}$Universidad de Santiago de Compostela, Santiago de Compostela, Spain\\
$ ^{37}$European Organization for Nuclear Research (CERN), Geneva, Switzerland\\
% ...
$ ^{40}$Nikhef National Institute for Subatomic Physics, Amsterdam, The Netherlands\\
% ...
$ ^{45}$H.H. Wills Physics Laboratory, University of Bristol, Bristol, United Kingdom\\
% ...
$ ^{50}$School of Physics and Astronomy, University of Glasgow, Glasgow, United Kingdom\\
$ ^{51}$Oliver Lodge Laboratory, University of Liverpool, Liverpool, United Kingdom\\
% ...
}
\end{flushleft}

%% file: body.tex
Decays of $B$ mesons to three-body hadronic charmless final states provide an interesting environment to search for \CP violation through the study of its signatures in the Dalitz plot~\cite{Miranda1,*Miranda2}.  
Theoretical predictions are mostly based on quasi-two-body decays to intermediate states, \eg $\rho^0\Kpm$ and $\Kstarz(892)\pipm$ for \kpipi decays and $\phi \Kpm$ for \kkk decays (see, \eg~Ref.~\cite{Neubert}). 
These intermediate states are accessible through amplitude analyses of data,
such as those performed by the Belle and the BaBar collaborations, who reported evidence of \CP violation in the intermediate channel $\rho^0\Kpm$~\cite{bellek2pi,BaBark2pi} in \kpipi decays and more recently in the channel $\phi \Kpm$~\cite{BaBarkkk} in \kkk decays.
However, the inclusive \CP asymmetry of \kpipi and \kkk decays was found to be consistent with zero.

For direct \CP violation to occur, two interfering amplitudes with different weak and strong phases must be involved in the decay process~\cite{BSS1979}.  
Large \CP violation effects have been observed in charmless two-body $B$ meson decays such as $\Bz\to\Kpm\pimp$ and $\Bs\to\Kmp\pipm$~\cite{LHCb-PAPER-2013-018}. 
However, the source of the strong phase difference in these processes is not well understood, which limits the potential to use these measurements to search for physics beyond the Standard Model.  
One possible source of the required strong phase is from final-state hadron rescattering, which can occur between two or more decay channels with the same flavour quantum numbers, such as \kpipi and \kkk~\cite{Marshak, Wolfenstein, Branco, livro_Bigi}.  
This effect, referred to as ``compound \CP violation"~\cite{Soni2005} is constrained by \CPT conservation so that the sum of the partial decay widths, for all channels with the same final-state quantum numbers related by the S-matrix, must be equal for charge-conjugated decays.

In this Letter we report measurements of the inclusive \CP-violating asymmetries in \kpipi and \kkk decays with unprecedented precision. 
The inclusion of charge-conjugate decay modes is implied except in the asymmetry definitions. 
The \CP asymmetry in \Bpm decays to a final state $f^{\pm}$ is defined as
\begin{equation}
\acp(\Bpm \to f^{\pm}) = \Phi [ \Gamma(\Bm \to  f^{-}) , \Gamma(\Bp \to  f^{+})],
\end{equation}
where $\Phi[X,Y]\equiv (X-Y)/(X+Y)$ is the asymmetry operator, 
$\Gamma$ is the decay width, 
and the final states are $f^{\pm}=\Kpm \pip\pim$ or $f^{\pm}=\Kpm \Kp\Km$.
We also study their asymmetry distributions across the phase space.

The \lhcb detector~\cite{Alves:2008zz} is a single-arm forward
spectrometer covering the \mbox{pseudorapidity} range $2<\eta <5$,
designed for the study of particles containing \bquark or \cquark
quarks. 
The analysis is based on $pp$ collision data, corresponding to an integrated luminosity of 1.0\,fb$^{-1}$, collected in 2011 at a centre-of-mass energy of 7~TeV. 

Events are selected by a trigger~\cite{LHCb-DP-2012-004} that consists of a hardware stage, based on information from the calorimeter and muon systems, followed by a software stage, which applies a full event reconstruction. 
Candidate events are first required to pass a hardware trigger, which selects particles with a large transverse energy. 
The software trigger requires a two-, three- or four-track secondary vertex with a high sum of the transverse momenta, \pt, of the tracks and a significant displacement from the primary $pp$ interaction vertices~(PVs). 
At least one track should have $\pt > 1.7\gevc$ and $\chisq_{\rm IP}$ with respect to any primary interaction greater than 16, where $\chisq_{\rm IP}$ is defined as the difference between the \chisq of a given PV reconstructed with and without the considered track. 
A multivariate algorithm is used for the identification of secondary vertices consistent with the decay of a \bquark hadron.

A set of selection criteria is applied to reconstruct $B$ mesons and suppress the combinatorial backgrounds. 
The \Bpm decay products are required to satisfy a set of selection criteria on their momenta, transverse momenta, the $\chisq_{\rm IP}$ of the final-state tracks, and the distance of closest approach between any two tracks. 
The $B$ candidates are required to have $\pt > 1.7\gevc$, $\chisq_{\rm IP}<10$ and displacement from any PV greater than 3~mm. 
Additional requirements are applied to variables related to the $B$ meson production and decay, 
such as quality of the track fits for the decay products, and the angle between the $B$ candidate momentum and the direction of flight from the primary vertex to the decay vertex. 
Final-state kaons and pions are further selected using particle identification information, provided by two ring-imaging Cherenkov detectors~\cite{LHCb-DP-2012-003}. 
The kinematic selection is common to both decay channels, while the particle identification selection is specific to each final state. 
Charm contributions are removed by excluding the regions of $\pm 30 \mevcc$ around the $\Dz$ mass in the two-body invariant masses \mpipi, \mkpi and \mkk. 
The contribution of the \jpsik decay is also excluded from the \kpipi sample by removing the mass region $3.05<\mpipi<3.15\gevcc$. 

\begin{figure*}[tb]
\centering
\includegraphics[width=0.49\linewidth]{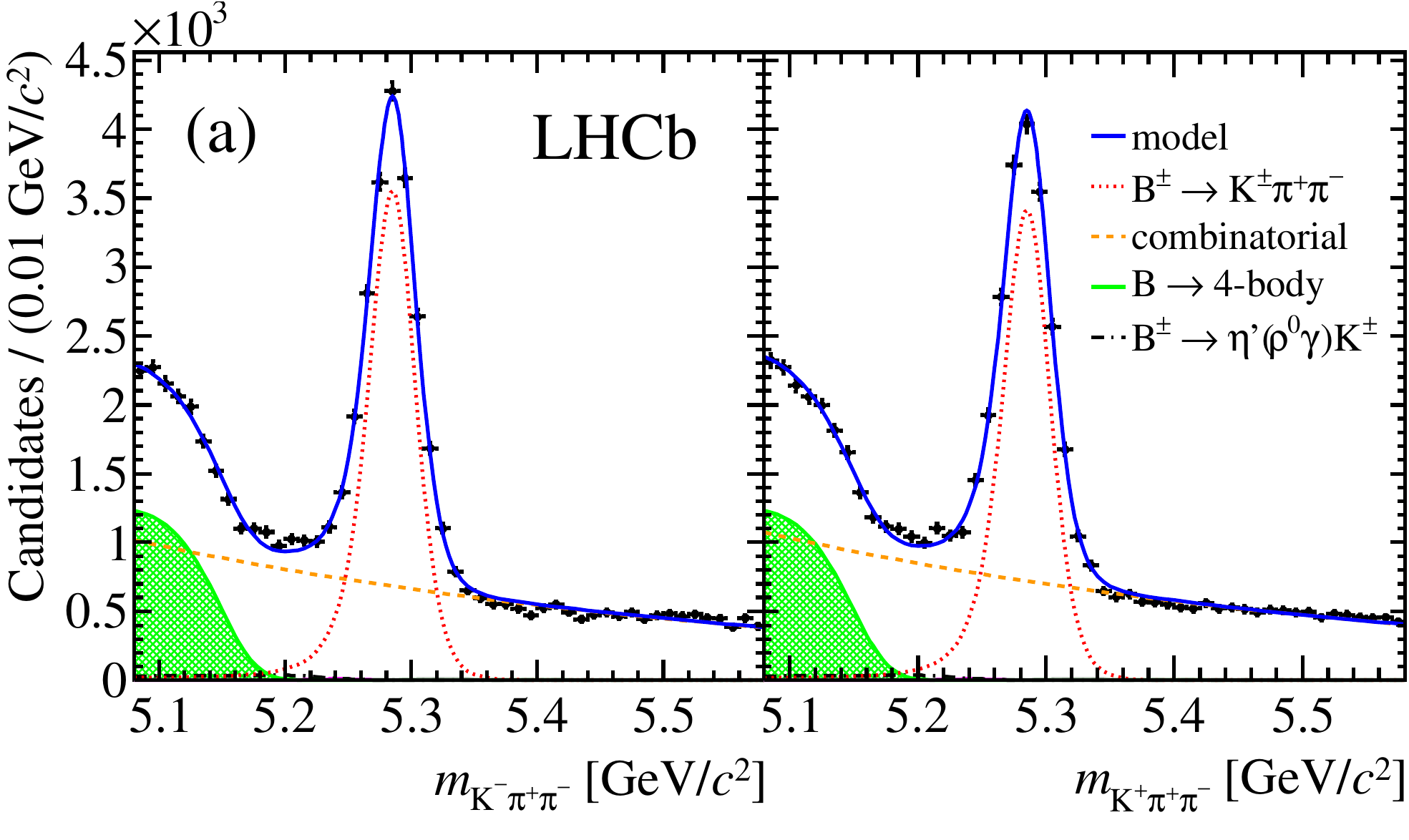}
\includegraphics[width=0.49\linewidth]{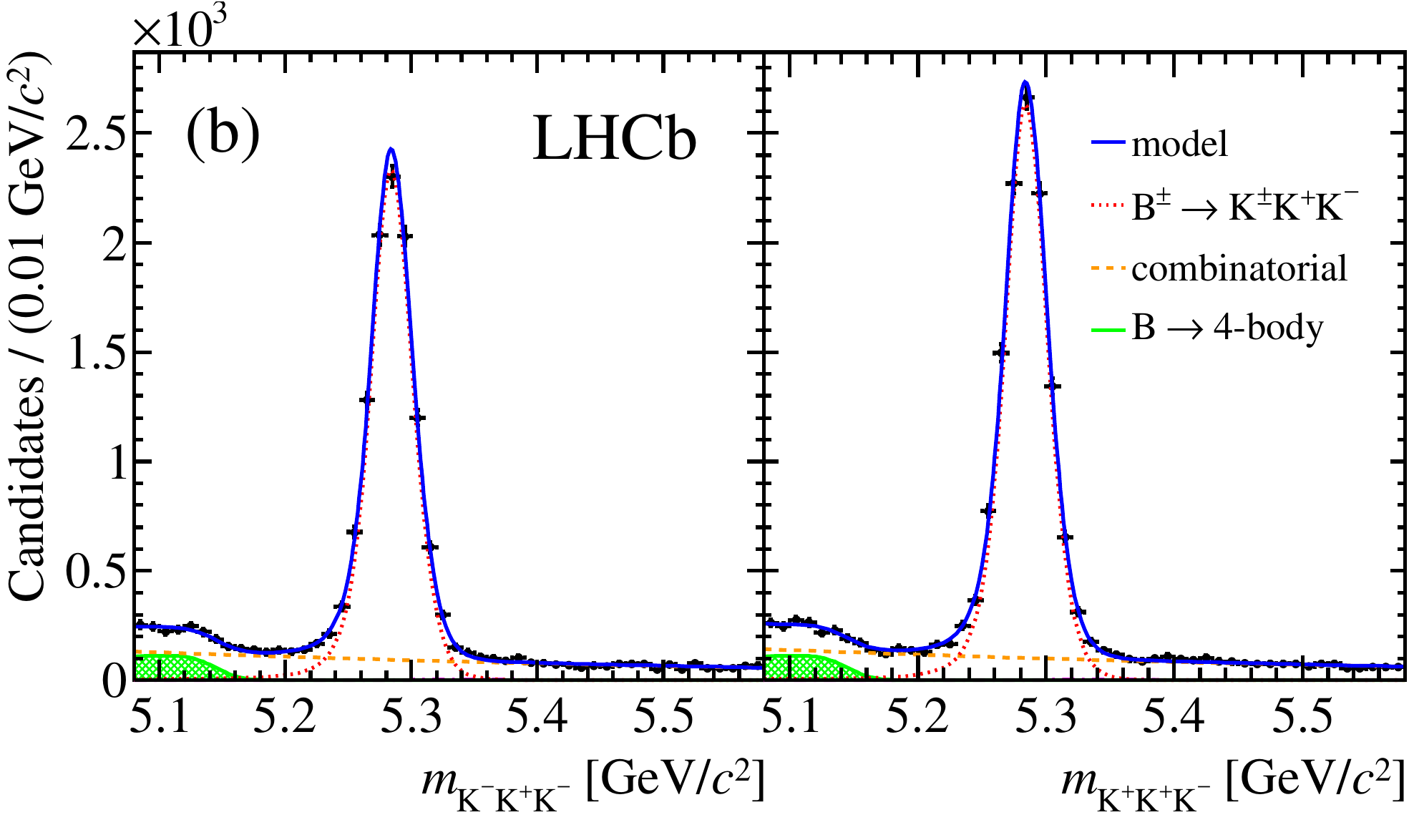}
\caption{Invariant mass spectra of (a) \kpipi decays and (b) \kkk decays. The left panel in each figure shows the \Bm modes and the right panel shows the \Bp modes.
The results of the unbinned maximum likelihood fits are overlaid. The main components of the fit are also shown. }
\label{MassFit}
\end{figure*}

The simulated events used in this analysis are generated 
using \pythia~6.4~\cite{Sjostrand:2006za} with a specific \lhcb configuration~\cite{LHCb-PROC-2010-056}.  
Decays of hadronic particles are produced by \evtgen~\cite{Lange:2001uf}, in which final-state radiation is generated using \photos~\cite{Golonka:2005pn}. 
The interaction of the generated particles with the detector and its response are implemented using the \geant toolkit~\cite{Allison:2006ve, *Agostinelli:2002hh} as described in Ref.~\cite{LHCb-PROC-2011-006}.

Unbinned extended maximum likelihood fits to the mass spectra of the selected \Bpm candidates are performed. 
The \kpipi and \kkk signal components are parameterised by so-called {\mbox{Cruijff}} functions~\cite{Cruijff} to account for the asymmetric effect of final-state radiation on the signal shape. 
The combinatorial background is described by an exponential function, and the background due to partially reconstructed four-body $B$ decays is parameterised by an ARGUS function~\cite{Argus} convolved with a Gaussian resolution function. 
Peaking backgrounds occur due to decay modes with one misidentified particle, and consist of the channels \kkpi, \pipipi and $B^{\pm} \to \eta'(\rho^0 \gamma) K^{\pm}$ for the \kpipi mode, and \kkpi for the \kkk mode. 
The shapes and yields of the peaking backgrounds  are obtained from simulation.
The invariant mass spectra of the \kpipi and \kkk candidates are shown in Fig.~\ref{MassFit}.

The mass fits of the two samples are used to obtain the signal yields, $N(K\pi\pi)=35\,901\pm327$ and $N(K\!K\!K)=22\,119\pm 164$, and the raw asymmetries, $\acpraw(K\pi\pi)=0.020\pm 0.007$ and $\acpraw(K\!K\!K) = -0.060\pm 0.007$, where the uncertainties are statistical. 
In order to determine the \CP asymmetries, the measured raw asymmetries are corrected for effects induced by the detector acceptance and interactions of final-state particles with matter, as well as for a possible $\B$-meson production asymmetry. 
The \CP asymmetry is expressed in terms of the raw asymmetry and a correction $A_{\Delta}$,
\begin{equation}
\!\!\! \acp \!=\! \acpraw \! -\! A_{\Delta} , \qquad
A_{\Delta}\! =\! \adetk \!+\! \aprod .
\label{eq:acpsum}
\end{equation}
Here \adetk is the kaon detection asymmetry, given in terms of the charge-conjugate kaon detection efficiencies $\varepsilon_D$ by $\adetk = \Phi [\varepsilon_D(\Km), \varepsilon_D(\Kp)]$, 
and \aprod is the production asymmetry, defined from the \Bpm production rates, $R(\Bpm)$, as $\aprod = \Phi [R(\Bm), R(\Bp)]$.
The decay products are regarded as a pair of charge-conjugate hadrons $h^+h^-=\pip\pim,\Kp\Km$, and a kaon with the same charge as the \Bpm meson, whose detection asymmetry is given by \adetk.

The correction term $A_{\Delta}$ is measured from data using a sample of approximately $6.3 \times 10^4$ $\Bpm \to \jpsi (\mup\mu^-)\Kpm$ decays. 
The \jpsik sample passes the same trigger, kinematic, and kaon particle identification selection as the signal samples, and it has a similar event topology. 
The kaons from \jpsik decay also have similar kinematics in the laboratory frame to those from the \kpipi and \kkk modes.  
The correction is obtained from the raw asymmetry of the \jpsik mode as 
\begin{equation}
A_{\Delta} = \acpraw(\jpsi K) - \acp(\jpsi K)  , 
\label{deltaJpsik}
\end{equation}
using the world average of the \CP asymmetry  $\acp(\jpsi K) = (0.1\pm 0.7)\%$~\cite{PDG2012}.
The \CP asymmetries of the \kpipi and \kkk channels are then determined using Eqs.~(\ref{eq:acpsum}) and~(\ref{deltaJpsik}).

Since the detector efficiencies for the signal modes are not flat in the Dalitz plot, and the raw asymmetries are also not uniformly distributed, an acceptance correction is applied to the integrated raw asymmetries. 
Furthermore, the detector acceptance and reconstruction depend on the trigger selection. 
The efficiency of the hadronic hardware trigger is found to have a small charge asymmetry for final-state kaons. 
Therefore, the data are divided into two 
samples with respect to the hadronic hardware trigger decision: events with candidates selected by the hadronic trigger, and events selected by other triggers independently of the signal candidate. 
In order to apply Eq.~(\ref{deltaJpsik}) to \kkk events selected by the hadronic hardware trigger, the difference in trigger efficiencies caused by the presence of three kaons compared to one kaon is taken into account. 
An acceptance correction is applied to each trigger sample of the \kpipim and \kkkm modes. 
It is determined by the ratio between the \Bm and \Bp average efficiencies in simulated events, reweighted to reproduce the population in the Dalitz plot of signal data. 
The subtraction of $A_{\Delta}$ is performed separately for each trigger configuration.
The integrated \CP asymmetries are then the weighted averages of the \CP asymmetries for the two trigger samples.

The systematic uncertainties on the asymmetries are related to the mass fit models, possible trigger asymmetry, and phase-space acceptance correction. 
In order to estimate the uncertainty due to the choice of the signal mass shape, the initial model is replaced with the sum of a Gaussian and a Crystal Ball function~\cite{Skwarnicki:1986xj}. 
The uncertainty associated with the combinatorial background model is estimated by repeating the fit with a first-order polynomial. 
We evaluate three uncertainties related to the peaking backgrounds: one due to the uncertainty on their yields, another due to the difference in mass resolution between simulation and data, and a third due to their possible non-zero asymmetries. 
The largest deviations from the nominal results are accounted for as systematic uncertainties. 
The systematic uncertainties related to the possible asymmetry induced by the trigger selection are of two kinds: one due to an asymmetric response of the hadronic hardware trigger to kaons, and a second due to the choice of sample division by trigger decision. 
The former is evaluated by reweighting the \jpsik mode with the charge-separated kaon efficiencies from calibration data. 
The latter is determined by varying the trigger composition of the samples in order to estimate the systematic differences in trigger admixture between the signal channels and the \jpsik mode. 
Two distinct uncertainties are attributed to the phase-space acceptance corrections: one is obtained from the uncertainty on the detection efficiency given by the simulation, and the other is evaluated by varying the binning of the acceptance map. 
The systematic uncertainties for the measurements of $\acp(\kpipi)$ and $\acp(\kkk)$ are summarised in Table~\ref{tab:syst}.

\begin{table}[t]
    \small
  \caption{    \small
   Systematic uncertainties on $\acp(\kpipis)$ and $\acp(\kkks)$. The total systematic uncertainties are the sum in quadrature of the individual contributions.}
\begin{center}\begin{tabular}{lcc}
\hline \hline
   Systematic uncertainty                    & $\acp(K\pi\pi)$ & $\acp(K\!K\!K)$      \\ 
    \hline
Signal model       & 0.0010 & \;\;\;\,0.0002 \\
Combinatorial background  & 0.0006 & $<0.0001$  \\
Peaking background           & 0.0007 & \;\;\;\,0.0001   \\
Trigger asymmetry           & 0.0036 & \;\;\;\,0.0019   \\
Acceptance correction     & 0.0012 & \;\;\;\,0.0019   \\ 
\hline
Total     & 0.0040 & \;\;\;\,0.0027  \\
\hline \hline
  \end{tabular}\end{center}
\label{tab:syst}
\end{table}

The results obtained for the inclusive \CP asymmetries of the \kpipi and \kkk decays are
\begin{eqnarray}
 \acp( \kpipi)   &=&  0.032 \pm  0.008  \pm  0.004 \pm 0.007  ,  \nonumber  \\ [2mm]
\acp(\kkk)    &=&   -0.043 \pm 0.009  \pm  0.003 \pm 0.007  , \nonumber
\end{eqnarray}
where the first uncertainty is statistical, the second is the experimental systematic, and the third is due to the \CP asymmetry of the \jpsik reference mode~\cite{PDG2012}. 
The significances of the inclusive charge asymmetries, calculated by dividing the central values by the sum in quadrature of the statistical and both systematic uncertainties, are 2.8 standard deviations~($\sigma$) for \kpipi and $3.7\sigma$ for \kkk decays.

\begin{figure*}[tb]
\centering
\includegraphics[width=0.47\linewidth]{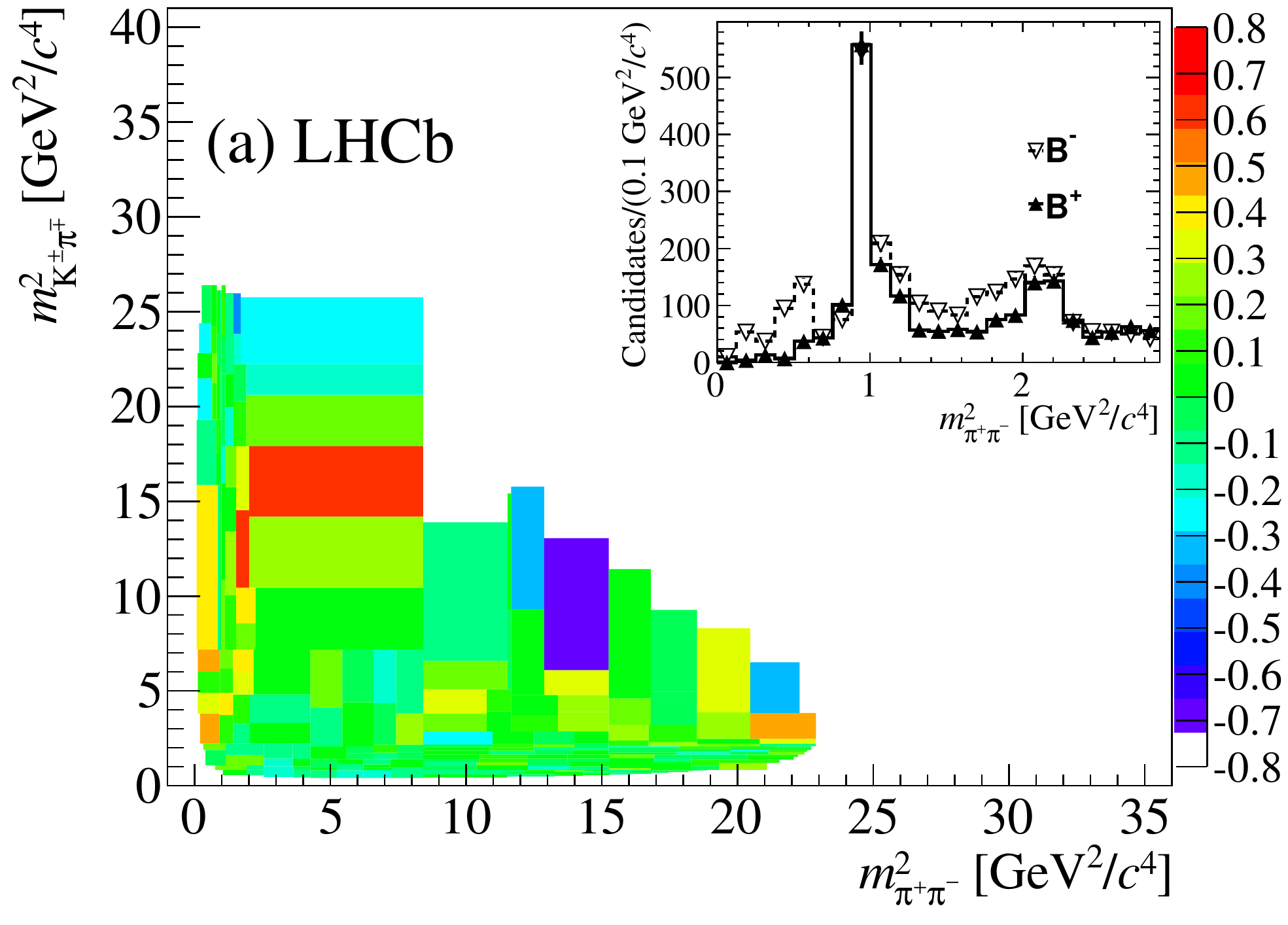}
\includegraphics[width=0.47\linewidth]{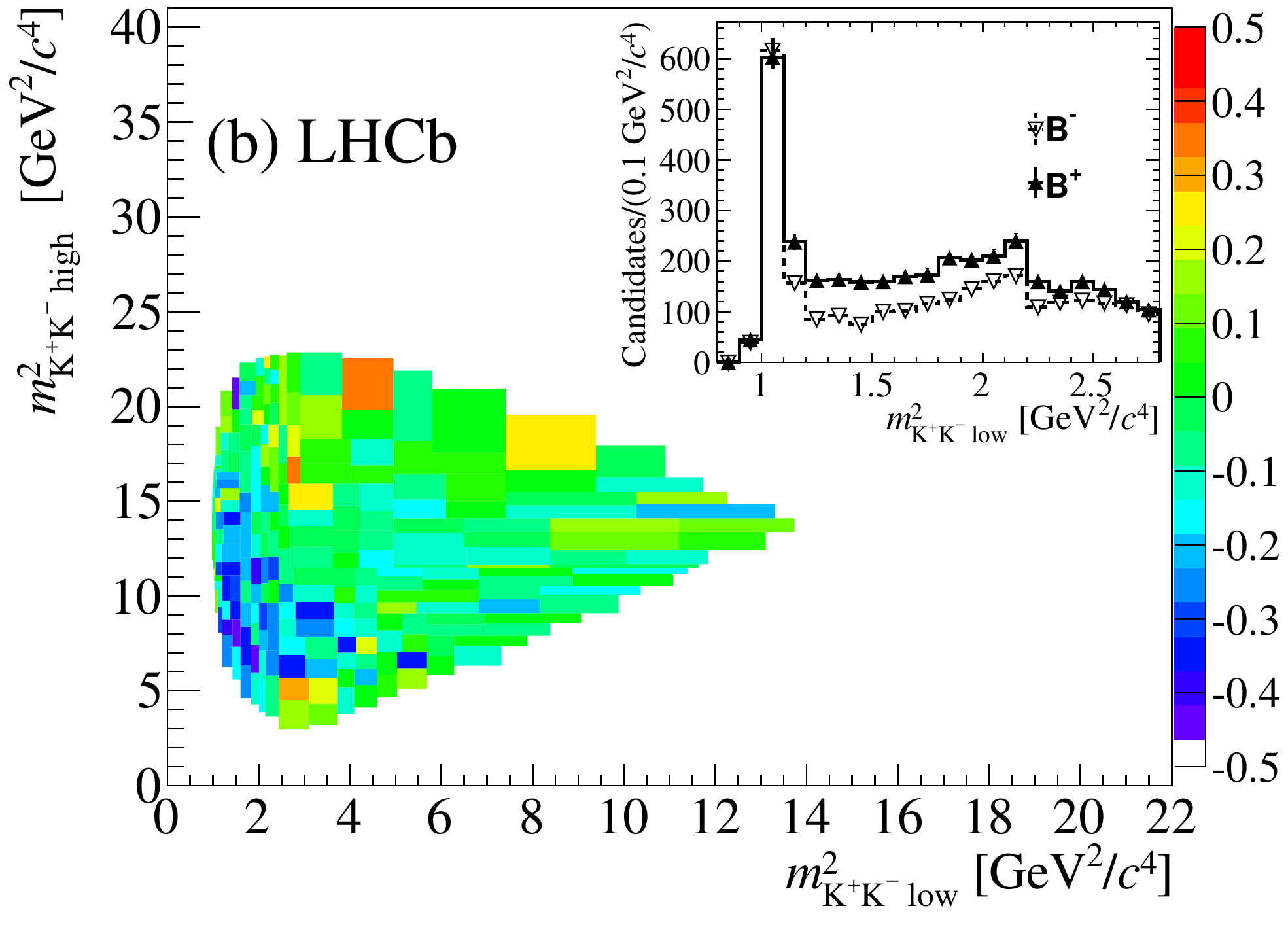}
\caption{
Asymmetries of the number of signal events in bins of the Dalitz plot, \acpn, for (a) \kpipi and  (b)~\kkk decays.  
The inset figures show the projections of the number of background-subtracted events in bins of  (left) the \mmpipi variable for $\mmkpi<15\gevgevcccc$ and (right) the \mmkklow variable for $\mmkkhi<15\gevgevcccc$. The distributions are not corrected for acceptance.
}
\label{Mirandizing}
\end{figure*}

\begin{figure*}[tb]
\centering
\includegraphics[width=0.49\linewidth]{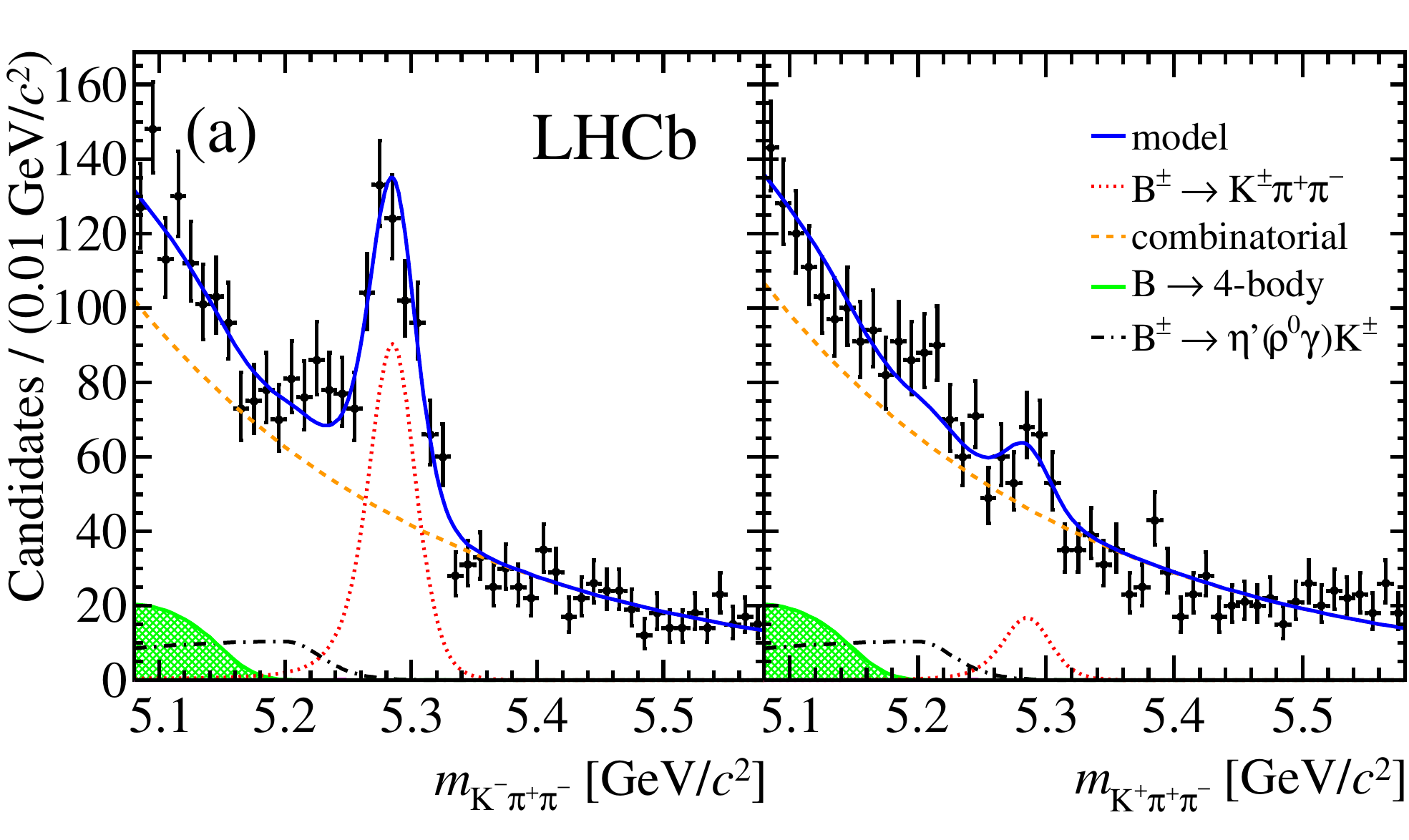}
\includegraphics[width=0.49\linewidth]{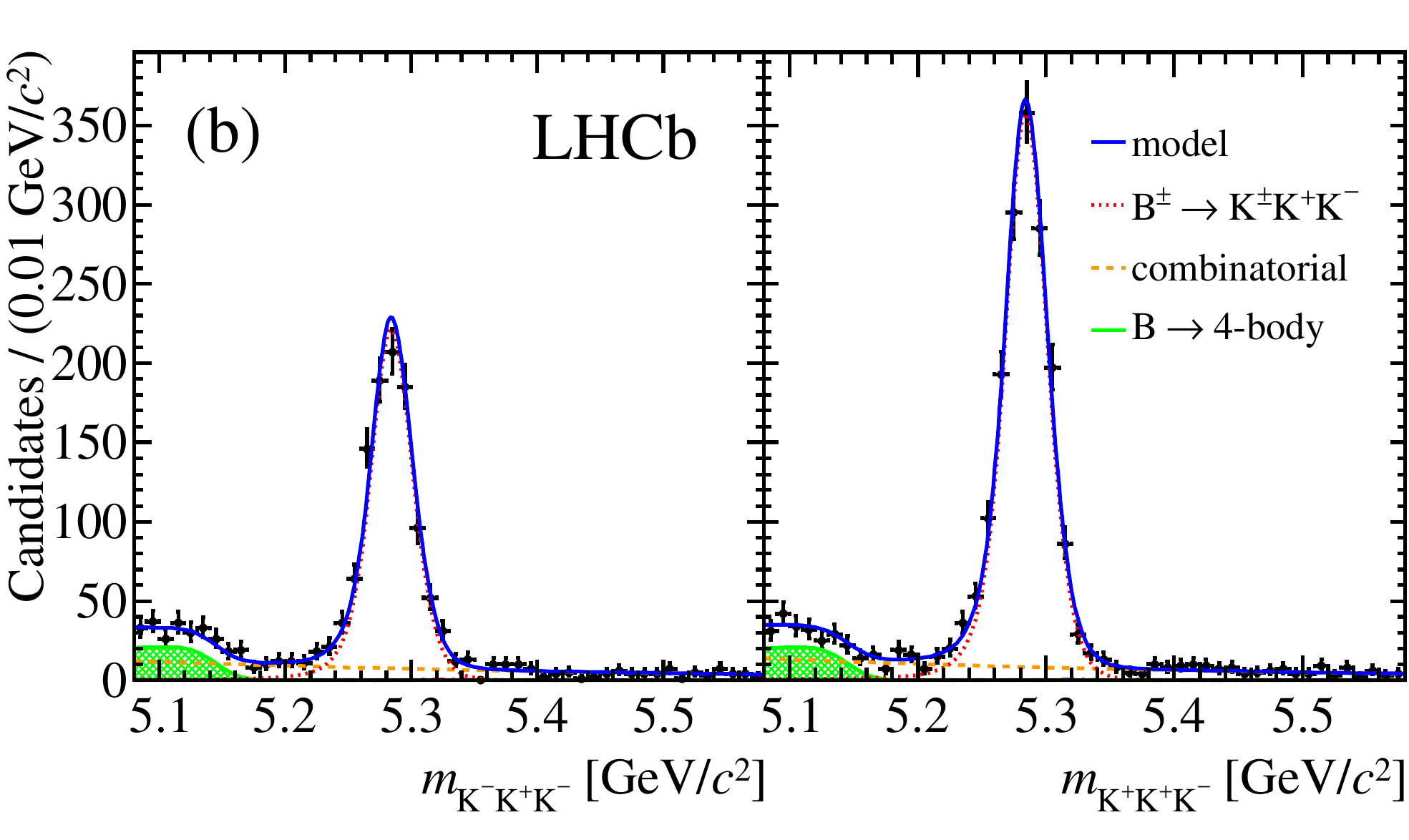}
\caption{Invariant mass spectra of (a) \kpipi decays in the region $0.08 < \mmpipi < 0.66\gevgevcccc$ and $\mmkpi < 15\gevgevcccc$, and  (b) \kkk decays in the region $1.2 < \mmkklow < 2.0\gevgevcccc$ and $\mmkkhi < 15\gevgevcccc$. The left panel in each figure shows the \Bm modes and the right panel shows the \Bp modes. The results of the unbinned maximum likelihood fits are overlaid. 
}
\label{MassFitRegion}
\end{figure*}

In addition to the inclusive charge asymmetries, we also study the asymmetry distributions in the two-dimensional phase space of two-body invariant masses. 
The background-subtracted Dalitz plot distributions of the signal region, defined as the region within three Gaussian widths from the signal peak, are divided into bins with equal numbers of events in the combined \Bm and \Bp samples. 
An asymmetry variable, $\acpn =  \Phi [N(\Bm), N(\Bp)]$, is computed from the number $N(\Bpm)$ of negative and positive entries in each bin of the background-subtracted Dalitz plots. 

The distributions of the \acpn variable in the Dalitz plots of \kpipi and \kkk are shown in Fig.~\ref{Mirandizing}, where the \kkk Dalitz plot is symmetrised and its two-body invariant mass squared variables are defined as $\mmkklow < \mmkkhi$. 
For \kpipi we identify a positive asymmetry located in the low $\pip\pim$ invariant mass region, around the $\rho(770)^0$ resonance, as seen by  Belle~\cite{bellek2pi} and BaBar~\cite{BaBark2pi}, and above the $f_0(980)$ resonance. 
This can be seen also in the inset figure of the $\pi^+\pi^-$ invariant mass projection, where there is an excess of \Bm candidates. 
No significant asymmetry is present in the low-mass region of the ${\Kpm \pimp}$ invariant mass projection.
The \acpn distribution of the \kkk mode reveals an asymmetry concentrated at low values of \mmkklow and \mmkkhi in the Dalitz plot. 
The distribution of the projection of the number of events onto the \mmkklow invariant mass (inset in the right plot of Fig.~\ref{Mirandizing}) shows that this asymmetry is not related to the $\phi(1020)$ resonance, but is instead located in the region $1.2 < \mmkklow < 2.0\gevgevcccc$. 

The \CP asymmetries are measured in two regions of phase space with large asymmetry. 
The \kkk region, $\mmkkhi < 15\gevgevcccc$ and $1.2 < \mmkklow < 2.0\gevgevcccc$, is defined such that the $\phi(1020)$ resonance is excluded. 
For the \kpipi mode we measure the \CP asymmetry of the region $\mmkpi < 15\gevgevcccc$ and $0.08 < \mmpipi < 0.66\gevgevcccc$, which spans the lowest $\pip\pim$ masses including the $\rho(770)^0$ resonance. 
Unbinned extended maximum likelihood fits are performed to the mass spectra of the candidates in the two regions, using the same models as the global fits. 
The spectra are shown in Fig.~\ref{MassFitRegion}. 
The resulting signal yields and raw asymmetries for the two regions are ${N^{\mathrm {reg}}(K\pi\pi)=552\pm47}$ and ${\acpraw^{\mathrm {reg}}(K\pi\pi)=0.687\pm0.078}$ for the \kpipi mode, and ${N^{\mathrm {reg}}(K\!K\!K)=2581\pm55}$ and ${\acpraw^{\mathrm {reg}}(K\!K\!K)=-0.239\pm0.020}$ for the \kkk mode. 
The \CP asymmetries are obtained from the raw asymmetries by applying an acceptance correction and subtracting the detection and production asymmetry correction $A_{\Delta}$ from \jpsik decays. 
The validity of the global $A_{\Delta}$ from \jpsik decays for the results in the regions was tested by comparing the kinematic distributions of their decay products.
Systematic uncertainties are estimated due to the signal models, trigger asymmetry, acceptance correction for the region and due to the limited validity of Eq.~(\ref{eq:acpsum}) for large asymmetries. 
The local charge asymmetries for the two regions are measured to be
\begin{eqnarray}
 \acp^{\mathrm {reg}}( K\pi\pi)   &=& 0.678 \pm 0.078 \pm 0.032 \pm 0.007 ,  \nonumber  \\ [2mm]
\acp^{\mathrm {reg}}(K\!K\!K)   &=&  -0.226  \pm 0.020 \pm 0.004 \pm 0.007  , \nonumber
\end{eqnarray}
where the first uncertainty is statistical, the second is the experimental systematic, and the third is due to the \CP asymmetry of the \jpsik reference mode.

In conclusion, we have measured the inclusive \CP asymmetries of the \kpipi and \kkk modes with significances of $2.8\sigma$ and $3.7\sigma$, respectively. 
The latter represents the first evidence of an inclusive \CP asymmetry in charmless three-body $B$ decays. 
These charge asymmetries are not uniformly distributed in the phase space. 
For  \kpipi decays, we observe positive asymmetries at low $\pip\pim$ masses, around the $\rho(770)^0$ resonance as indicated by Belle~\cite{bellek2pi}  and BaBar~\cite{BaBark2pi}, and also above the $f_0(980)$ resonance, where it is not clearly associated to resonances.   
Although it is possible to identify the signature of the $\rho(770)^0$ resonance for any value of \mmkpi, 
the asymmetry appears only at low $\Kpm\pimp$ mass around the $\rho(770)^0$ invariant mass. 
A signature of \CP violation is present in the \kkk Dalitz plot, mostly concentrated in the region of low \mmkklow and low \mmkkhi.
A similar pattern of the \CP asymmetry was shown in the preliminary results of the \kkpi and \pipipi decay modes by LHCb~\cite{LHCb-CONF-2012-028}, in which the positive asymmetries are at low $\pip\pim$ masses and the negative at low $\Kp\Km$ masses, both not clearly associated to intermediate resonant states.

Moreover, the excess of events in the \kpipim with respect to the \kpipip sample is comparable to the excess of \kkkp with respect to the \kkkm mode. 
This apparent correlation, together with the inhomogeneous \CP asymmetry distribution in the Dalitz plot, could be related to compound \CP violation. 
Since the \kpipi and \kkk modes have the same flavour quantum numbers (as do the pair \kkpi and \pipipi), 
\CP violation induced by hadron rescattering could play an important role in these charmless three-body $B$ decays. 
In order to quantify a possible compound \CP asymmetry, the introduction of new amplitude analysis techniques, which would take into account the presence of hadron rescattering in three-body $B$ decays, is necessary.

%% file: acknowledgements.tex
\section*{Acknowledgements}

\noindent We express our gratitude to our colleagues in the CERN
accelerator departments for the excellent performance of the LHC. We
thank the technical and administrative staff at the LHCb
institutes. We acknowledge support from CERN and from the national
agencies: CAPES, CNPq, FAPERJ and FINEP (Brazil); NSFC (China);
CNRS/IN2P3 and Region Auvergne (France); BMBF, DFG, HGF and MPG
(Germany); SFI (Ireland); INFN (Italy); FOM and NWO (The Netherlands);
SCSR (Poland); ANCS/IFA (Romania); MinES, Rosatom, RFBR and NRC
``Kurchatov Institute'' (Russia); MinECo, XuntaGal and GENCAT (Spain);
SNSF and SER (Switzerland); NAS Ukraine (Ukraine); STFC (United
Kingdom); NSF (USA). We also acknowledge the support received from the
ERC under FP7. The Tier1 computing centres are supported by IN2P3
(France), KIT and BMBF (Germany), INFN (Italy), NWO and SURF (The
Netherlands), PIC (Spain), GridPP (United Kingdom). We are thankful
for the computing resources put at our disposal by Yandex LLC
(Russia), as well as to the communities behind the multiple open
source software packages that we depend on.

%% file: main.bbl
\ifx\mcitethebibliography\mciteundefinedmacro
\PackageError{LHCb.bst}{mciteplus.sty has not been loaded}
{This bibstyle requires the use of the mciteplus package.}\fi
\providecommand{\href}[2]{#2}
\begin{mcitethebibliography}{10}
\mciteSetBstSublistMode{n}
\mciteSetBstMaxWidthForm{subitem}{\alph{mcitesubitemcount})}
\mciteSetBstSublistLabelBeginEnd{\mcitemaxwidthsubitemform\space}
{\relax}{\relax}

\bibitem{Miranda1}
I.~Bediaga {\em et~al.}, \ifthenelse{\boolean{articletitles}}{{\it {On a CP
  anisotropy measurement in the Dalitz plot}},
  }{}\href{http://dx.doi.org/10.1103/PhysRevD.80.096006}{Phys.\ Rev.\  {\bf
  D80} (2009) 096006}, \href{http://arxiv.org/abs/0905.4233}{{\tt
  arXiv:0905.4233}}\relax
\mciteBstWouldAddEndPuncttrue
\mciteSetBstMidEndSepPunct{\mcitedefaultmidpunct}
{\mcitedefaultendpunct}{\mcitedefaultseppunct}\relax
\EndOfBibitem
\bibitem{Miranda2}
I.~Bediaga {\em et~al.}, \ifthenelse{\boolean{articletitles}}{{\it {Second
  generation of `Miranda procedure' for CP violation in Dalitz studies of $B$
  (and $D$ and $\tau$) decays}},
  }{}\href{http://dx.doi.org/10.1103/PhysRevD.86.036005}{Phys.\ Rev.\  {\bf
  D86} (2012) 036005}, \href{http://arxiv.org/abs/1205.3036}{{\tt
  arXiv:1205.3036}}\relax
\mciteBstWouldAddEndPuncttrue
\mciteSetBstMidEndSepPunct{\mcitedefaultmidpunct}
{\mcitedefaultendpunct}{\mcitedefaultseppunct}\relax
\EndOfBibitem
\bibitem{Neubert}
M.~Beneke and M.~Neubert, \ifthenelse{\boolean{articletitles}}{{\it {QCD
  factorization for $B \to PP$ and $B \to PV$ decays}},
  }{}\href{http://dx.doi.org/10.1016/j.nuclphysb.2003.09.026}{Nucl.\ Phys.\
  {\bf B675} (2003) 333}, \href{http://arxiv.org/abs/hep-ph/0308039}{{\tt
  arXiv:hep-ph/0308039}}\relax
\mciteBstWouldAddEndPuncttrue
\mciteSetBstMidEndSepPunct{\mcitedefaultmidpunct}
{\mcitedefaultendpunct}{\mcitedefaultseppunct}\relax
\EndOfBibitem
\bibitem{bellek2pi}
Belle collaboration, A.~Garmash {\em et~al.},
  \ifthenelse{\boolean{articletitles}}{{\it {Evidence for large direct CP
  violation in $B^{\pm} \to \rho^0(770) K^{\pm}$ from analysis of the
  three-body charmless $B^{\pm} \to K^{\pm} \pi^{\pm} \pi^{\mp}$ decay}},
  }{}\href{http://dx.doi.org/10.1103/PhysRevLett.96.251803}{Phys.\ Rev.\ Lett.\
   {\bf 96} (2006) 251803}, \href{http://arxiv.org/abs/hep-ex/0512066}{{\tt
  arXiv:hep-ex/0512066}}\relax
\mciteBstWouldAddEndPuncttrue
\mciteSetBstMidEndSepPunct{\mcitedefaultmidpunct}
{\mcitedefaultendpunct}{\mcitedefaultseppunct}\relax
\EndOfBibitem
\bibitem{BaBark2pi}
BaBar collaboration, B.~Aubert {\em et~al.},
  \ifthenelse{\boolean{articletitles}}{{\it {Evidence for direct CP violation
  from Dalitz-plot analysis of $B^\pm \to K^\pm \pi^\mp \pi^\pm$}},
  }{}\href{http://dx.doi.org/10.1103/PhysRevD.78.012004}{Phys.\ Rev.\ D {\bf
  78} (2008) 012004}, \href{http://arxiv.org/abs/0803.4451}{{\tt
  arXiv:0803.4451}}\relax
\mciteBstWouldAddEndPuncttrue
\mciteSetBstMidEndSepPunct{\mcitedefaultmidpunct}
{\mcitedefaultendpunct}{\mcitedefaultseppunct}\relax
\EndOfBibitem
\bibitem{BaBarkkk}
BaBar collaboration, J.-P. Lees {\em et~al.},
  \ifthenelse{\boolean{articletitles}}{{\it {Study of CP violation in
  Dalitz-plot analyses of $B^0 \to K^+K^-K^0_{S}$, $B^+ \to K^+K^-K^+$, and
  $B^+ \to K^0_{S}K^0_{S}K^+$}},
  }{}\href{http://dx.doi.org/10.1103/PhysRevD.85.112010}{Phys.\ Rev.\ D {\bf
  85} (2012) 112010}, \href{http://arxiv.org/abs/1201.5897}{{\tt
  arXiv:1201.5897}}\relax
\mciteBstWouldAddEndPuncttrue
\mciteSetBstMidEndSepPunct{\mcitedefaultmidpunct}
{\mcitedefaultendpunct}{\mcitedefaultseppunct}\relax
\EndOfBibitem
\bibitem{BSS1979}
M.~Bander, D.~Silverman, and A.~Soni, \ifthenelse{\boolean{articletitles}}{{\it
  {\CP noninvariance in the decays of heavy charged quark systems}},
  }{}\href{http://dx.doi.org/10.1103/PhysRevLett.43.242}{Phys.\ Rev.\ Lett.\
  {\bf 43} (1979) 242}\relax
\mciteBstWouldAddEndPuncttrue
\mciteSetBstMidEndSepPunct{\mcitedefaultmidpunct}
{\mcitedefaultendpunct}{\mcitedefaultseppunct}\relax
\EndOfBibitem
\bibitem{LHCb-PAPER-2013-018}
LHCb collaboration, R.~Aaij {\em et~al.},
  \ifthenelse{\boolean{articletitles}}{{\it {First observation of \CP violation
  in the decays of $B^0_s$ mesons}},
  }{}\href{http://dx.doi.org/0.1103/PhysRevLett.110.221601}{Phys.\ Rev.\ Lett.\
   {\bf 110} (2013) 221601}, \href{http://arxiv.org/abs/1304.6173}{{\tt
  arXiv:1304.6173}}\relax
\mciteBstWouldAddEndPuncttrue
\mciteSetBstMidEndSepPunct{\mcitedefaultmidpunct}
{\mcitedefaultendpunct}{\mcitedefaultseppunct}\relax
\EndOfBibitem
\bibitem{Marshak}
R.~Marshak, Riazuddin, and C.~Ryan, {\em {Theory of weak interactions in
  particle physics}}, Wiley-Interscience, New York, NY, USA, 1969\relax
\mciteBstWouldAddEndPuncttrue
\mciteSetBstMidEndSepPunct{\mcitedefaultmidpunct}
{\mcitedefaultendpunct}{\mcitedefaultseppunct}\relax
\EndOfBibitem
\bibitem{Wolfenstein}
L.~Wolfenstein, \ifthenelse{\boolean{articletitles}}{{\it {Final state
  interactions and CP violation in weak decays}},
  }{}\href{http://dx.doi.org/10.1103/PhysRevD.43.151}{Phys.\ Rev.\  {\bf D43}
  (1991) 151}\relax
\mciteBstWouldAddEndPuncttrue
\mciteSetBstMidEndSepPunct{\mcitedefaultmidpunct}
{\mcitedefaultendpunct}{\mcitedefaultseppunct}\relax
\EndOfBibitem
\bibitem{Branco}
G.~C. Branco, L.~Lavoura, and J.~P. Silva,
  \ifthenelse{\boolean{articletitles}}{{\it {CP violation}}, }{}Int.\ Ser.\
  Monogr.\ Phys.\  {\bf 103} (1999) 1\relax
\mciteBstWouldAddEndPuncttrue
\mciteSetBstMidEndSepPunct{\mcitedefaultmidpunct}
{\mcitedefaultendpunct}{\mcitedefaultseppunct}\relax
\EndOfBibitem
\bibitem{livro_Bigi}
I.~I. Bigi and A.~Sanda, \ifthenelse{\boolean{articletitles}}{{\it {CP
  violation}}, }{}Camb.\ Monogr.\ Part.\ Phys.\ Nucl.\ Phys.\ Cosmol.\  {\bf 9}
  (2000) 1\relax
\mciteBstWouldAddEndPuncttrue
\mciteSetBstMidEndSepPunct{\mcitedefaultmidpunct}
{\mcitedefaultendpunct}{\mcitedefaultseppunct}\relax
\EndOfBibitem
\bibitem{Soni2005}
H.-Y. Cheng, C.-K. Chua, and A.~Soni, \ifthenelse{\boolean{articletitles}}{{\it
  {Final state interactions in hadronic $B$ decays}},
  }{}\href{http://dx.doi.org/10.1103/PhysRevD.71.014030}{Phys.\ Rev.\  {\bf
  D71} (2005) 014030}, \href{http://arxiv.org/abs/hep-ph/0409317}{{\tt
  arXiv:hep-ph/0409317}}\relax
\mciteBstWouldAddEndPuncttrue
\mciteSetBstMidEndSepPunct{\mcitedefaultmidpunct}
{\mcitedefaultendpunct}{\mcitedefaultseppunct}\relax
\EndOfBibitem
\bibitem{Alves:2008zz}
LHCb collaboration, A.~A. Alves~Jr. {\em et~al.},
  \ifthenelse{\boolean{articletitles}}{{\it {The \lhcb detector at the LHC}},
  }{}\href{http://dx.doi.org/10.1088/1748-0221/3/08/S08005}{JINST {\bf 3}
  (2008) S08005}\relax
\mciteBstWouldAddEndPuncttrue
\mciteSetBstMidEndSepPunct{\mcitedefaultmidpunct}
{\mcitedefaultendpunct}{\mcitedefaultseppunct}\relax
\EndOfBibitem
\bibitem{LHCb-DP-2012-004}
R.~Aaij {\em et~al.}, \ifthenelse{\boolean{articletitles}}{{\it {The \lhcb
  trigger and its performance in 2011}},
  }{}\href{http://dx.doi.org/10.1088/1748-0221/8/04/P04022}{JINST {\bf 8}
  (2013) P04022}, \href{http://arxiv.org/abs/1211.3055}{{\tt
  arXiv:1211.3055}}\relax
\mciteBstWouldAddEndPuncttrue
\mciteSetBstMidEndSepPunct{\mcitedefaultmidpunct}
{\mcitedefaultendpunct}{\mcitedefaultseppunct}\relax
\EndOfBibitem
\bibitem{LHCb-DP-2012-003}
M.~Adinolfi {\em et~al.}, \ifthenelse{\boolean{articletitles}}{{\it
  {Performance of the \lhcb RICH detector at the LHC}},
  }{}\href{http://dx.doi.org/10.1140/epjc/s10052-013-2431-9}{Eur.\ Phys.\ J.\
  {\bf C73} (2013) 2431}, \href{http://arxiv.org/abs/1211.6759}{{\tt
  arXiv:1211.6759}}\relax
\mciteBstWouldAddEndPuncttrue
\mciteSetBstMidEndSepPunct{\mcitedefaultmidpunct}
{\mcitedefaultendpunct}{\mcitedefaultseppunct}\relax
\EndOfBibitem
\bibitem{Sjostrand:2006za}
T.~Sj\"{o}strand, S.~Mrenna, and P.~Skands,
  \ifthenelse{\boolean{articletitles}}{{\it {PYTHIA 6.4 physics and manual}},
  }{}\href{http://dx.doi.org/10.1088/1126-6708/2006/05/026}{JHEP {\bf 05}
  (2006) 026}, \href{http://arxiv.org/abs/hep-ph/0603175}{{\tt
  arXiv:hep-ph/0603175}}\relax
\mciteBstWouldAddEndPuncttrue
\mciteSetBstMidEndSepPunct{\mcitedefaultmidpunct}
{\mcitedefaultendpunct}{\mcitedefaultseppunct}\relax
\EndOfBibitem
\bibitem{LHCb-PROC-2010-056}
I.~Belyaev {\em et~al.}, \ifthenelse{\boolean{articletitles}}{{\it {Handling of
  the generation of primary events in \gauss, the \lhcb simulation framework}},
  }{}\href{http://dx.doi.org/10.1109/NSSMIC.2010.5873949}{Nuclear Science
  Symposium Conference Record (NSS/MIC) {\bf IEEE} (2010) 1155}\relax
\mciteBstWouldAddEndPuncttrue
\mciteSetBstMidEndSepPunct{\mcitedefaultmidpunct}
{\mcitedefaultendpunct}{\mcitedefaultseppunct}\relax
\EndOfBibitem
\bibitem{Lange:2001uf}
D.~J. Lange, \ifthenelse{\boolean{articletitles}}{{\it {The EvtGen particle
  decay simulation package}},
  }{}\href{http://dx.doi.org/10.1016/S0168-9002(01)00089-4}{Nucl.\ Instrum.\
  Meth.\  {\bf A462} (2001) 152}\relax
\mciteBstWouldAddEndPuncttrue
\mciteSetBstMidEndSepPunct{\mcitedefaultmidpunct}
{\mcitedefaultendpunct}{\mcitedefaultseppunct}\relax
\EndOfBibitem
\bibitem{Golonka:2005pn}
P.~Golonka and Z.~Was, \ifthenelse{\boolean{articletitles}}{{\it {PHOTOS Monte
  Carlo: a precision tool for QED corrections in $Z$ and $W$ decays}},
  }{}\href{http://dx.doi.org/10.1140/epjc/s2005-02396-4}{Eur.\ Phys.\ J.\  {\bf
  C45} (2006) 97}, \href{http://arxiv.org/abs/hep-ph/0506026}{{\tt
  arXiv:hep-ph/0506026}}\relax
\mciteBstWouldAddEndPuncttrue
\mciteSetBstMidEndSepPunct{\mcitedefaultmidpunct}
{\mcitedefaultendpunct}{\mcitedefaultseppunct}\relax
\EndOfBibitem
\bibitem{Allison:2006ve}
GEANT4 collaboration, J.~Allison {\em et~al.},
  \ifthenelse{\boolean{articletitles}}{{\it {Geant4 developments and
  applications}}, }{}\href{http://dx.doi.org/10.1109/TNS.2006.869826}{IEEE
  Trans.\ Nucl.\ Sci.\  {\bf 53} (2006) 270}\relax
\mciteBstWouldAddEndPuncttrue
\mciteSetBstMidEndSepPunct{\mcitedefaultmidpunct}
{\mcitedefaultendpunct}{\mcitedefaultseppunct}\relax
\EndOfBibitem
\bibitem{Agostinelli:2002hh}
GEANT4 collaboration, S.~Agostinelli {\em et~al.},
  \ifthenelse{\boolean{articletitles}}{{\it {GEANT4: a simulation toolkit}},
  }{}\href{http://dx.doi.org/10.1016/S0168-9002(03)01368-8}{Nucl.\ Instrum.\
  Meth.\  {\bf A506} (2003) 250}\relax
\mciteBstWouldAddEndPuncttrue
\mciteSetBstMidEndSepPunct{\mcitedefaultmidpunct}
{\mcitedefaultendpunct}{\mcitedefaultseppunct}\relax
\EndOfBibitem
\bibitem{LHCb-PROC-2011-006}
M.~Clemencic {\em et~al.}, \ifthenelse{\boolean{articletitles}}{{\it {The \lhcb
  simulation application, \gauss: design, evolution and experience}},
  }{}\href{http://dx.doi.org/10.1088/1742-6596/331/3/032023}{{J.\ Phys.\ \!\!:
  Conf.\ Ser.\ } {\bf 331} (2011) 032023}\relax
\mciteBstWouldAddEndPuncttrue
\mciteSetBstMidEndSepPunct{\mcitedefaultmidpunct}
{\mcitedefaultendpunct}{\mcitedefaultseppunct}\relax
\EndOfBibitem
\bibitem{Cruijff}
BaBar collaboration, P.~del Amo~Sanchez {\em et~al.},
  \ifthenelse{\boolean{articletitles}}{{\it {Study of $B \to X\gamma$ decays
  and determination of $|V_{td}/V_{ts}|$}},
  }{}\href{http://dx.doi.org/10.1103/PhysRevD.82.051101}{Phys.\ Rev.\  {\bf
  D82} (2010) 051101}, \href{http://arxiv.org/abs/1005.4087}{{\tt
  arXiv:1005.4087}}\relax
\mciteBstWouldAddEndPuncttrue
\mciteSetBstMidEndSepPunct{\mcitedefaultmidpunct}
{\mcitedefaultendpunct}{\mcitedefaultseppunct}\relax
\EndOfBibitem
\bibitem{Argus}
ARGUS collaboration, H.~Albrecht {\em et~al.},
  \ifthenelse{\boolean{articletitles}}{{\it {Search for $b\to s\gamma$ in
  exclusive decays of $B$ mesons}},
  }{}\href{http://dx.doi.org/10.1016/0370-2693(89)91177-5}{Phys.\ Lett.\  {\bf
  B229} (1989) 304}\relax
\mciteBstWouldAddEndPuncttrue
\mciteSetBstMidEndSepPunct{\mcitedefaultmidpunct}
{\mcitedefaultendpunct}{\mcitedefaultseppunct}\relax
\EndOfBibitem
\bibitem{PDG2012}
Particle Data Group, J.~Beringer {\em et~al.},
  \ifthenelse{\boolean{articletitles}}{{\it {\href{http://pdg.lbl.gov/}{Review
  of particle physics}}},
  }{}\href{http://dx.doi.org/10.1103/PhysRevD.86.010001}{Phys.\ Rev.\  {\bf
  D86} (2012) 010001}\relax
\mciteBstWouldAddEndPuncttrue
\mciteSetBstMidEndSepPunct{\mcitedefaultmidpunct}
{\mcitedefaultendpunct}{\mcitedefaultseppunct}\relax
\EndOfBibitem
\bibitem{Skwarnicki:1986xj}
T.~Skwarnicki, {\em {A study of the radiative cascade transitions between the
  Upsilon-prime and Upsilon resonances}}, PhD thesis, Institute of Nuclear
  Physics, Krakow, 1986,
  {\href{http://inspirehep.net/record/230779/files/230779.pdf}{DESY-F31-86-02}%
}\relax
\mciteBstWouldAddEndPuncttrue
\mciteSetBstMidEndSepPunct{\mcitedefaultmidpunct}
{\mcitedefaultendpunct}{\mcitedefaultseppunct}\relax
\EndOfBibitem
\bibitem{LHCb-CONF-2012-028}
{LHCb collaboration}, \ifthenelse{\boolean{articletitles}}{{\it {Evidence for
  \CP violation in $B^{\pm} \to K^+K^-\pi^{\pm}$ and $B^{\pm} \to
  \pi^+\pi^-\pi^{\pm}$ decays}}, }{}
  \href{http://cdsweb.cern.ch/search?p={LHCb-CONF-2012-028}&f=reportnumber&act%
ion_search=Search&c=LHCb+Reports&c=LHCb+Conference+Proceedings&c=LHCb+Conferen%
ce+Contributions&c=LHCb+Notes&c=LHCb+Theses&c=LHCb+Papers}
  {{LHCb-CONF-2012-028}}\relax
\mciteBstWouldAddEndPuncttrue
\mciteSetBstMidEndSepPunct{\mcitedefaultmidpunct}
{\mcitedefaultendpunct}{\mcitedefaultseppunct}\relax
\EndOfBibitem
\end{mcitethebibliography}
